\documentclass[final,1p,sort&compress,times]{elsarticle}

\usepackage{amssymb}
\usepackage{hyperref}
\usepackage{bm}
\usepackage{graphicx}

\journal{Physics Letters A}

\begin{document}

\begin{frontmatter}



\title{Reconstructing the relaxation dynamics induced by an unknown heat bath}


\author{Jochen Rau}
\ead{jochen.rau@q-info.org}
\ead[url]{www.q-info.org}

\address{Institut f\"ur Theoretische Physik,
Johann Wolfgang Goethe-Universit\"at,
Max-von-Laue-Str. 1, 
60438 Frankfurt am Main,
Germany}

\begin{abstract}
In quantum state tomography, one potential source of error is uncontrolled contact of the system with a heat bath whose detailed properties are not known,
and whose impact on the system moreover varies between different runs of the experiment.
Precisely these variations provide a handle for reconstructing the system's effective relaxation dynamics.
I propose a pertinent estimation scheme which is based on a steepest-descent ansatz and maximum likelihood.
After reconstructing the relaxation dynamics,
the original quantum state of the system can be constrained to a curve in state space.
\end{abstract}

\begin{keyword}

quantum state tomography
\sep
heat bath
\sep
relaxation dynamics
\sep
maximum likelihood

\PACS
03.65.Wj
\sep
02.50.Tt


\end{keyword}

\end{frontmatter}



\section{\label{intro}Motivation}

Quantum state tomography \cite{book:stateestimation} is prone to errors,
of various origins.
For instance,
samples might be small,
the observables measured might not be informationally complete,
or
measurement devices might be inaccurate.
In this letter, I  focus on one further source of systematic error:
The quantum system might be coupled in an uncontrollable fashion to a heat bath of fixed but unknown temperature (say, a piece of equipment or the surrounding air), whose effect may moreover 
vary randomly from sample to sample
(say, because the duration of interaction varies between different runs of the experiment).
Transient contact with the bath then triggers partial relaxation (to varying extent) of the system towards thermal equilibrium.
In general, neither
the original quantum state of the system nor
the pertinent system-bath dynamics are known;
and so neither is
the target equilibrium state,
let alone the trajectory in state space leading from the original state towards the latter.
Under such adverse circumstances,
how can one hope to learn anything about the original quantum state?
In this letter,
I  provide a simple, approximate scheme that allows one at least to constrain the unknown quantum state to a one-dimensional submanifold of state space,
based on the outcomes of multiple independent runs of the experiment.
The reconstruction of this one-dimensional constraint manifold assumes a generic relaxation dynamics and uses the maximum likelihood approximation.

\section{\label{relaxation}Generic relaxation dynamics}

In this section 
I  first discuss the generic relaxation dynamics close to equilibrium
and subsequently propose an extension to states further away.
Close to equilibrium, in the linear response regime,
states typically relax towards equilibrium exponentially,
\begin{equation}
	\rho(t)-\sigma 
	=
	\exp(-t/\tau_{\rm rel}) [\rho(0)-\sigma]
	,
\end{equation}
on some pertinent time scale $\tau_{\rm rel}$. 
Here $\sigma$ denotes the equilibrium state of the system.
In the setting considered here, both $\tau_{\rm rel}$ and $\sigma$ are fixed but unknown.
The relaxation time $\tau_{\rm rel}$ should be larger than the typical contact times with the bath so that relaxation of the system towards equilibrium is only partial.

Such exponential relaxation can be regarded as
a \textit{steepest-descent flow} of the relative entropy $S(\rho\|\sigma)$, in the following sense.
Given any set of observables $\{F_b\}$ which is informationally complete,
every state $\rho$ can be written in the Gibbs form
\begin{equation}
	\rho = Z(\lambda)^{-1} \exp\left[ (\ln\sigma-\langle\ln\sigma\rangle_\sigma)-\sum_b \lambda^{b} F_b \right]
	,
\label{gibbs}
\end{equation}
with properly adjusted Lagrange parameters $\{\lambda^{b}\}$
and partition function $Z(\lambda)$ \cite{ruskai:minrent,olivares+paris,PhysRevA.84.012101}.
(This is not to be confused with a thermal state:
A thermal state has the Gibbs form, too,
yet with only one observable (Hamiltonian) and Lagrange parameter (inverse temperature) in the exponent,
and with $\sigma$ replaced by the totally mixed state.)
The Lagrange parameters $\{\lambda^{b}\}$ and the expectation values $\{f_b:=\langle F_b\rangle_\rho\}$,
respectively,
constitute two possible choices for the coordinates in state space.
Associated with these coordinates are the respective basis vectors $\{\partial_b:=\partial/\partial\lambda^b\}$
and $\{\partial^b:=\partial/\partial f_b\}$, 
as well as their one-form duals $\{d\lambda^a\}$ and $\{df_a\}$,
related via
$d\lambda^a(\partial_b)=df_b(\partial^a)=\delta^a_b$ \cite{schutz:book}.
The transformation from one coordinate basis to the other is effected by the correlation matrix
$C_{ab}:=\partial^2 \ln Z(\lambda)/\partial\lambda^a\partial\lambda^b$,
\begin{equation}
	\partial_a= -C_{ab} \partial^b
	\ ,\ 
	\partial^a = -(C^{-1})^{ab} \partial_b
\end{equation}
(where I have adopted the convention that identical upper and lower indices are to be summed over),
and likewise the transformation of their duals,
\begin{equation}
	df_a = -C_{ab} d\lambda^b
	\ ,\ 
	d\lambda^a = -(C^{-1})^{ab} df_b
	.
\end{equation}
Being symmetric and positive, the correlation matrix also provides a natural Riemannian metric on state space,
the Bogoliubov-Kubo-Mori (BKM) metric \cite{bengtsson:book,balian:physrep,10.1063/1.530611,10.1063/1.531535,Jochen200083,grasselli:uniqueness}.
Its elements may be regarded as the components of the metric tensor
\begin{equation}
	C:=C_{ab} d\lambda^a \otimes d\lambda^b
	.
\end{equation}
The relative entropy $S(\rho\|\sigma)$ has the gradient \cite{PhysRevA.84.012101}
\begin{equation}
	dS(\rho\|\sigma) = - \lambda^a df_a
	,
\end{equation}
and so the pertinent steepest-descent curve (in the BKM metric introduced above) has a tangent vector 
\begin{equation}
	V \propto C^{-1}(-dS(\rho\|\sigma)) = -\lambda^a \partial_a
	.
\end{equation}
Consequently, on a steepest-descent trajectory the Lagrange parameters must evolve according to
$d\lambda^b/dt \propto (-\lambda^a \partial_a)\lambda^b = -\lambda^b$
and thus relax exponentially towards $\lambda=0$, which corresponds to the state $\sigma$.
In the linear response regime,
exponential relaxation of the Lagrange parameters in turn implies exponential relaxation of expectation values,
and so indeed of the statistical operator,
Q.E.D.

Identifying thus the generic relaxation dynamics near equilibrium with a steepest-descent flow  
opens the possibility to formulate a generic dynamics even in regions further away:
In the absence of detailed information about the system-bath dynamics,
and regardless of how far the state of the system might be from equilibrium,
I  henceforth assume that
the generic effect of a bath on the system is to move its state by some (unknown) distance along the steepest-descent trajectory towards some (unknown) equilibrium state $\sigma$.
Such an extension of the steepest-descent paradigm to full state space has been studied before,
and in the special case of the classical Fokker-Planck equation (with a particular choice of metric) 
has  been shown
to reproduce the effective dynamics excactly
\cite{Jordan1997265,10.1137/S0036141096303359}.
In the setting considered here, the steepest-descent algorithm induces an effective quantum operation on the state space of the system.
Close to equilibrium,
this effective quantum operation is a completely positive map.
Further away from equilibrium, however,
it might no longer be linear and thus not completely positive;
which suggests that,
in contrast to a completely positive map,
the steepest-descent ansatz may presume non-negligible initial correlations between system and bath.
Such initial correlations are not surprising in view of the fact that the properties of the heat bath are not known \textit{a priori};
they simply reflect the fact  
that learning something about the initial state of the system
will entail learning something about the  state of the bath, too.
(There are also other reasons why one should expect the effective quantum dynamics to be nonlinear,
see e.g. Ref. \cite{PhysRevA.82.052119}.)
It remains to be seen whether for a given steepest-descent flow there  exists always some compatible microscopic system-bath dynamics,
just as any completely positive map always arises from some underlying unitary evolution of the  (initially uncorrelated) composite system-bath state. 
I conjecture that this is the case, but defer its proof to future work.

Any steepest-descent trajectory towards $\sigma$ can  be parametrized in the form
\begin{equation}
	\rho(\gamma) \propto \exp\left[ (\ln\sigma-\langle\ln\sigma\rangle_\sigma)- \gamma G \right]
	,
\label{trajectory}
\end{equation}
with some \textit{single} observable $G$ and associated Lagrange parameter $\gamma$.
I prove this assertion in two steps:
(i)
any curve parametrized in this form is  a steepest-descent curve towards $\sigma$;
and
(ii)
for every initial state $\rho_0$ there exists a curve of this form passing through it.
The first statement becomes evident when one considers the above parametric form to be
a special case of the more general form (\ref{gibbs}),
with the pair $(\gamma,G)$ being one of the informationally complete $\{(\lambda^b,F_b)\}$ and
the Lagrange parameters associated with all $F_b\neq G$ equal to zero.
If these other Lagrange parameters are initially zero, they remain zero under the steepest-descent dynamics
$d\lambda^b/dt \propto -\lambda^b$.
So indeed, a curve parametrized as above describes a steepest descent towards $\sigma$.
As for the second statement,
for any initial $\rho_0$ (except for states on the boundary of state space) 
the choice $G=\ln\sigma-\ln\rho_0$ will yield a steepest-descent curve passing through both
$\rho_0$ (at $\gamma=1$) and $\sigma$ ($\gamma=0$).
Given $\rho_0$ and $\sigma$,
this choice for $G$ is unique up to multiplicative and additive constants.

Given the imperfect data from multiple runs of the experiment that have been affected
(to varying degree) by uncontrolled contact of the system with the same unknown heat bath,
a precise reconstruction of the original quantum state is clearly not possible.
However, assuming that transient contact with the bath triggers a generic relaxation dynamics as described above,
it is at least possible to constrain the original quantum state to some steepest-descent trajectory.
Having established the parametric form (\ref{trajectory}) of such a trajectory,
the  task is then to infer from the available experimental data the observable $G$
(up to multiplicative and additive constants)
as well as \textit{some} state (not necessarily the target state $\sigma$) which lies on this trajectory.
This is the inference task to which I turn in the next section.

\section{\label{theory}Inferring the constraint curve}

The experiment is run multiple times.
In the $i$th run, after the supposed disturbance by the bath,
one performs complete quantum state tomography on a sample of size $N_i$ and obtains the tomographic image $\mu_i$.
Different runs of the experiment are disturbed  by the same bath but to varying extent (say, because the contact time varies between runs).
As a result,
the original quantum state is shifted on its relaxation trajectory towards equilibrium to varying degree;
and so rather than being clustered  
around a single point (which would yield a unique state estimate),
the images $\{\mu_i\}$ are expected to be 
spread out along this trajectory.

Assuming that the relaxation trajectory has the steepest-descent form (\ref{trajectory}),
the problem of inferring the observable $G$ from images that correspond to various values of $\gamma$
is analogous to the problem of estimating a Hamiltonian from thermal data at various temperatures.
Assuming that the tomographic images are not too far apart, and
expanding $G$ in terms of the informationally complete set $\{F_b\}$,
\begin{equation}
	G=-\xi^b F_b
\end{equation}
(with the same index summation convention as above),
the  log-likelihood of observing the data $\{\mu_i\}$ is asymptotically
(i.e., for sufficiently large sample sizes) 
given by
\begin{equation}
	L(\{\mu_i\}|\xi,\sigma)
	\sim
	(N/2)
	[\langle\Gamma\rangle_{\xi} - (C^{-1})^{ab}\delta f_a(\xi,\sigma) \delta f_b(\xi,\sigma)]
	,
\label{xi_likelihood}
\end{equation}
modulo additive terms that do not depend on $\xi$ or $\sigma$ \cite{PhysRevA.84.052101}.
Here 
$N:=\sum_i N_i$,
and
\begin{equation}
	\langle \Gamma\rangle_\xi
	:=
	\frac{\Gamma_{ab}\xi^a \xi^b}{C_{cd}\xi^c \xi^d}
\end{equation}
is the ``expectation value'' of the covariance matrix 
\begin{equation}
	\Gamma_{ab}:=\sum_{i} w_i (f_a^i-\bar{f}_a)(f_b^i-\bar{f}_b)
	.
\end{equation}
The latter in turn depends on the deviations of the 
sample means $f_b^i:=\langle F_b\rangle_{\mu_i}$
from their weighted averages $\bar{f}_b := \sum_i w_i f_b^i$,
with the $i$th sample carrying relative weight $w_i:=N_i/N$.
The correlation matrix $C_{ab}$ is evaluated at the center of mass
$\bar{\mu}:=\sum_i w_i \mu_i$
of the tomographic images.
Finally, the variations $\delta f$ are given by
\begin{equation}
	\delta f_b(\xi,\sigma):=\langle F_b\rangle_{\bar{\pi}(\xi,\sigma)}-\bar{f}_b
	,
\end{equation}
where $\bar{\pi}(\xi,\sigma)$ is the unique state of the form (\ref{trajectory}) 
which yields 
$\langle G\rangle_{\bar{\pi}}=\langle G\rangle_{\bar{\mu}}$.

Unlike in the problem of estimating a Hamiltonian from thermal data where $\sigma$ corresponds to a given reference state, 
the target state $\sigma$ is not fixed \textit{a priori}
but is itself a variable to be inferred.
To maximize the above log-likelihood as a function of both $\xi$ and $\sigma$,
the steepest-descent trajectory  must  maximize $\langle \Gamma\rangle_\xi$ as well as render $\delta f=0$.
The first requirement implies that   
$\vec{\xi}$ must be the dominant eigenvector (i.e., the eigenvector associated with the largest eigenvalue) of $\bm{C}^{-1}\bm{\Gamma}$,
the product of the inverse correlation and covariance matrices.
Such an eigenvector condition is a familiar result 
in principal component analysis \cite{roweis:em,roweis:unify,bishop:bayesian_pca,bishop:variational_pca,RSSB:RSSB196}.
The second requirement, on the other hand, is met by any $\sigma$ that ensures $\bar{\pi}=\bar{\mu}$,
and hence by any steepest-descent trajectory that passes through the center of mass $\bar{\mu}$.
Thus one can infer from the experimental data both a maximum likelihood estimate for $\vec{\xi}$, and hence for $G$,
and a state (namely $\bar{\mu}$) through which the steepest-descent trajectory must pass. 
Together, these   characterize the steepest-descent trajectory uniquely.
The original quantum state of the system must lie somewhere on this trajectory.

\section{\label{toy}Example: qubits}

\begin{figure}[tbp]
\begin{center}
\includegraphics[width=9cm]{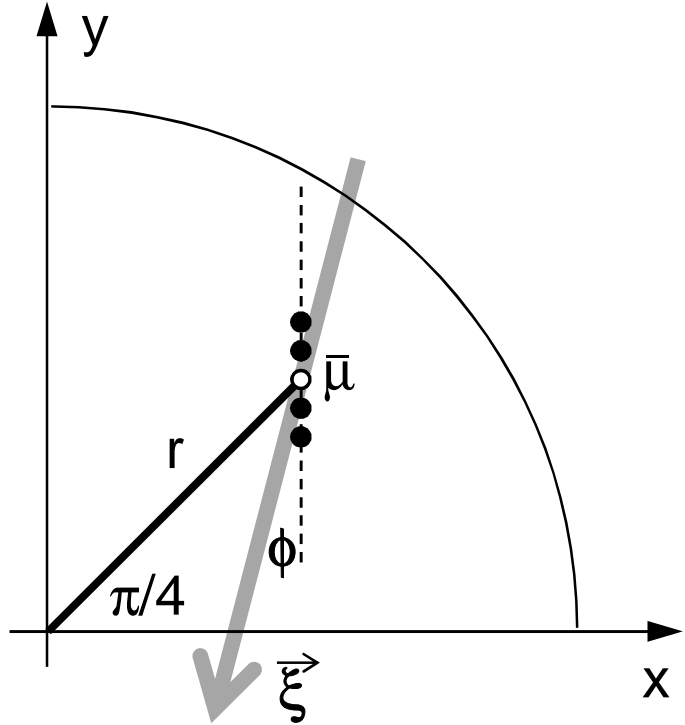}
\end{center}
\caption{First quadrant of a two-dimensional section ($z=0$) of the Bloch sphere. 
The black dots indicate the tomographic images $\{\mu_i\}$ obtained in different runs of the experiment, and the small circle their center of mass $\bar{\mu}$.
The Bloch vector associated with the center of mass  is assumed to have no $z$ component ($\langle\sigma_z\rangle_{\bar{\mu}}=0$),
azimuth $\pi/4$, and length $r$.
Even though the  images are aligned parallel to the $y$ axis,
the inferred direction $\vec{\xi}$ of the effective force which drives the relaxation towards equilibrium is tilted against that axis by an angle $\phi$.
This angle grows from $\phi=0$ at $r=0$ to $\phi=\pi/4$ at $r=1$.
}
\label{qubit}
\end{figure}

The simplest example 
is state tomography on an exchangeable sequence of qubits, emitted by some i.i.d. quantum source.
The tomographic experiment is performed several times,
with the $i$th run yielding the tomographic image $\mu_i$.
Rather than isotropically around a  point, these images $\{\mu_i\}$ turn out to be clustered along a curve in state space;
say, along a straight line parallel to the $y$ axis of the Bloch sphere (Fig. \ref{qubit}).
This suggests that in between runs of the experiment, 
there is some uncontrolled change in the environment which is characterized by a single parameter.
An obvious candidate is the coupling to an unknown heat bath,
with the contact time fluctuating in between runs.

For a qubit, 
the informationally complete set of observables $\{F_b\}$ can be taken to be the Pauli matrices.
The associated parameter vector $\vec{\xi}$ points in the direction of the effective
``dissipative force'' which drives the qubit towards equilibrium.
Due to the nontrivial geometry of quantum states this force is in general \textit{not} parallel
to the spatial orientation of the data cluster
(here: the $y$ axis) as one might na{\"i}vely expect.
Rather, it is tilted against this axis by some angle $\phi$.
This angle can be calculated as follows.
Assuming that the measured data are perfectly aligned parallel to the $y$ axis as shown in Fig. \ref{qubit},
all entries of the covariance matrix vanish except for $\Gamma_{yy}\neq 0$.
And for the center-of-mass state $\bar{\mu}$ located in the $x-y$ plane as shown in Fig. \ref{qubit},
the inverse correlation matrix has the form
\begin{equation}
	\bm{C}^{-1}(\bar{\mu})=
	\left(
	\begin{array}{ccc}
	C^{-1}_{xx} & C^{-1}_{xy} & 0 \\
	C^{-1}_{yx} & C^{-1}_{yy} & 0 \\
	0 & 0 & C^{-1}_{zz}
\end{array}
\right)
.
\end{equation}
The requirement that
$\vec{\xi}$ be the dominant eigenvector of $\bm{C}^{-1}\bm{\Gamma}$ then implies
$\xi_z=0$ and $\xi_x/\xi_y=C^{-1}_{xy}/C^{-1}_{yy}$.
As a consequence,  $\vec{\xi}$ is tilted against the $y$ axis by the angle
\begin{equation}
	\phi=\arctan(C^{-1}_{xy}/C^{-1}_{yy})
	.
\end{equation}

This tilting angle vanishes whenever the center of mass lies on one of the two ($x$ or $y$) axes.
Away from the axes, however, the tilting angle is non-zero, and increases as the center of mass moves closer to the surface of the Bloch sphere.
To illustrate the latter, I consider a center-of-mass state on the $x-y$ diagonal,
with azimuth $\pi/4$ and variable Bloch vector length $r$.
The relevant elements of the inverse correlation matrix are then 
$C^{-1}_{yy}=1/(1-r^2)$ and $C^{-1}_{xy}=C^{-1}_{yy}-(\tanh^{-1}r)/r$.
This yields a function $\phi(r)$ which increases monotonically
from $\phi(0)=0$ to $\phi(1)=\pi/4$,
and which to a reasonable degree of accuracy
can be approximated by $\phi(r)\approx(\pi/4)r^2$.
Only near the origin of the Bloch sphere, therefore, and hence for highly mixed states,
does the inferred direction of the effective dissipative force coincide with the ``na{\"i}ve'' estimate based on the spatial orientation of the data cluster.
For  states that are (nearly) pure,
on the other hand,
the reconstruction scheme  presented here may yield a direction $\vec{\xi}$ which differs significantly from that na{\"i}ve estimate.

\section{\label{discussion}Conclusions}

Whenever an i.i.d. sequence of quantum systems is subjected to disturbance by the same heat bath,
yet to an extent that varies randomly between samples,
multiple runs of  quantum state tomography are expected to yield data spread out along a curve in state space.
From the shape and location of this sprawl,
one can infer the system's effective relaxation dynamics 
under the influence of the bath.
Owing to the nontrivial geometry of quantum states,
the result of this inference can be rather counterintuitive.
In the qubit example, the inferred direction of the effective dissipative force generally deviated from the principal axis of the data cluster.

There are several ways to extend the results of the present letter,
which are left to future work.
(i)
So far the reconstruction scheme for the effective relaxation dynamics does not take into account the prior distribution $\mbox{prob}(\xi,\sigma)$ of its parameters.
Doing so will lead to a Bayesian modification of the maximum likelihood framework presented here.
A full Bayesian analysis should also include a study of the error bars on $\xi$ and $\sigma$.
(ii)
Many physical systems, especially larger ones,
exhibit not a single relaxation time but a whole hierarchy of time scales pertaining to the relaxation of different sets of degrees of freedom.
In this case thermalization occurs in stages,
on successively longer time scales \cite{Balian1987229,rau:physrep}.
Provided the typical contact times with the bath are shorter than the smallest scale in that hierarchy,
the reconstruction scheme presented here is still valid.
It pertains then to the first thermalization stage,
with $\sigma$ being no longer the equilibrium state but the target state of this first stage.
If contact times vary widely across different relaxation time scales, however,
the scheme must be adapted.
(iii)
The system might be disturbed by not just one but several baths, each to independently varying degree,
or the impact of a single bath might be governed by more than one parameter.
Then, rather than along a curve, tomographic images will be spread out in some higher-dimensional submanifold of state space. 
Inferring the dimensionality of this submanifold, as well as the associated parameters,
will require a further generalisation of the scheme presented here.
(iv)
Reconstructing the relaxation trajectory (or submanifold) is the first step towards the reconstruction of the system's {unperturbed} quantum state.
I argued that this unperturbed quantum state must lie somewhere on the relaxation trajectory (or submanifold);
but its precise location will depend on further assumptions, not considered in the present letter, 
about the distribution of the contact times and possibly other parameters of the bath.
(v)
Finally,
I consider it worthwhile to study in more detail the conceptual underpinning as well as possible modifications of the steepest-descent paradigm.
And on a speculative note,
taking the steepest-descent paradigm at face value and identifying (some suitable function of) relative entropy as actual ``time,''
one might even be tempted to try to establish a link to (equally speculative) ideas in other areas of physics that aim to reduce the notion of time to a distance between configurations \cite{barbour:timeless} or thermal properties \cite{0264-9381-11-12-007}.


\bibliographystyle{elsarticle-num}

\begin{thebibliography}{10}
\expandafter\ifx\csname url\endcsname\relax
  \def\url#1{\texttt{#1}}\fi
\expandafter\ifx\csname urlprefix\endcsname\relax\def\urlprefix{URL }\fi
\expandafter\ifx\csname href\endcsname\relax
  \def\href#1#2{#2} \def\path#1{#1}\fi

\bibitem{book:stateestimation}
M.~G.~A. Paris, J.~Reh{\'a}cek (Eds.), Vol. 649 of Lect. Notes Phys., Springer,
  2004.
\newblock \href {http://dx.doi.org/10.1007/b98673} {\path{doi:10.1007/b98673}}.

\bibitem{ruskai:minrent}
M.~B. Ruskai, Extremal properties of relative entropy in quantum statistical
  mechanics, Rep.\ Math.\ Phys. 26 (1988) 143.
\newblock \href {http://dx.doi.org/10.1016/0034-4877(88)90009-2}
  {\path{doi:10.1016/0034-4877(88)90009-2}}.

\bibitem{olivares+paris}
S.~Olivares, M.~G.~A. Paris, Quantum estimation via the minimum \mbox{Kullback}
  entropy principle, Phys. Rev. A 76 (2007) 042120.
\newblock \href {http://dx.doi.org/10.1103/PhysRevA.76.042120}
  {\path{doi:10.1103/PhysRevA.76.042120}}.

\bibitem{PhysRevA.84.012101}
J.~Rau, Inferring the \mbox{Gibbs} state of a small quantum system, Phys. Rev.
  A 84 (2011) 012101.
\newblock \href {http://dx.doi.org/10.1103/PhysRevA.84.012101}
  {\path{doi:10.1103/PhysRevA.84.012101}}.

\bibitem{schutz:book}
B.~F. Schutz, Geometrical methods of mathematical physics, Cambridge University
  Press, 1980.
\newblock \href {http://dx.doi.org/10.2277/0521298873}
  {\path{doi:10.2277/0521298873}}.

\bibitem{bengtsson:book}
I.~Bengtsson, K.~{\.{Z}}yczkowski, Geometry of quantum states, Cambridge
  University Press, 2006.
\newblock \href {http://dx.doi.org/10.2277/0521814510}
  {\path{doi:10.2277/0521814510}}.

\bibitem{balian:physrep}
R.~Balian, Y.~Alhassid, H.~Reinhardt, Dissipation in many-body systems: A
  geometric approach based on information theory, Phys.\ Rep. 131 (1986) 1.
\newblock \href {http://dx.doi.org/10.1016/0370-1573(86)90005-0}
  {\path{doi:10.1016/0370-1573(86)90005-0}}.

\bibitem{10.1063/1.530611}
D.~Petz, Geometry of canonical correlation on the state space of a quantum
  system, J. Math. Phys. 35 (1994) 780.
\newblock \href {http://dx.doi.org/10.1063/1.530611}
  {\path{doi:10.1063/1.530611}}.

\bibitem{10.1063/1.531535}
D.~Petz, C.~Sud{\'a}r, Geometries of quantum states, J. Math. Phys. 37 (1996)
  2662.
\newblock \href {http://dx.doi.org/10.1063/1.531535}
  {\path{doi:10.1063/1.531535}}.

\bibitem{Jochen200083}
J.~Dittmann, On the curvature of monotone metrics and a conjecture concerning
  the \mbox{Kubo-Mori} metric, Linear Algebra and its Applications 315 (2000)
  83.
\newblock \href {http://dx.doi.org/10.1016/S0024-3795(00)00130-0}
  {\path{doi:10.1016/S0024-3795(00)00130-0}}.

\bibitem{grasselli:uniqueness}
M.~Grasselli, R.~F. Streater, On the uniqueness of the \mbox{Chentsov} metric
  in quantum information geometry, Inf. Dim. Analysis, Quantum Prob. (IDAQP) 4
  (2001) 173.
\newblock \href {http://dx.doi.org/10.1142/S0219025701000462}
  {\path{doi:10.1142/S0219025701000462}}.

\bibitem{Jordan1997265}
R.~Jordan, D.~Kinderlehrer, F.~Otto, Free energy and the \mbox{Fokker-Planck}
  equation, Physica D 107 (1997) 265.
\newblock \href {http://dx.doi.org/10.1016/S0167-2789(97)00093-6}
  {\path{doi:10.1016/S0167-2789(97)00093-6}}.

\bibitem{10.1137/S0036141096303359}
R.~Jordan, D.~Kinderlehrer, F.~Otto, The variational formulation of the
  \mbox{Fokker-Planck} equation, SIAM J. Math. Anal. 29 (1998) 1.
\newblock \href {http://dx.doi.org/10.1137/S0036141096303359}
  {\path{doi:10.1137/S0036141096303359}}.

\bibitem{PhysRevA.82.052119}
H.~C. \"Ottinger, Nonlinear thermodynamic quantum master equation: Properties
  and examples, Phys. Rev. A 82 (2010) 052119.
\newblock \href {http://dx.doi.org/10.1103/PhysRevA.82.052119}
  {\path{doi:10.1103/PhysRevA.82.052119}}.

\bibitem{PhysRevA.84.052101}
J.~Rau, Assessing thermalization and estimating the \mbox{Hamiltonian} with
  output data only, Phys. Rev. A 84 (2011) 052101.
\newblock \href {http://dx.doi.org/10.1103/PhysRevA.84.052101}
  {\path{doi:10.1103/PhysRevA.84.052101}}.

\bibitem{roweis:em}
S.~T. Roweis, \mbox{EM} algorithms for \mbox{PCA} and \mbox{SPCA}, in: M.~I.
  Jordan, M.~J. Kearns, S.~A. Solla (Eds.), Advances in Neural Information
  Processing Systems 10, MIT Press, 1998, pp. 626--632.

\bibitem{roweis:unify}
S.~T. Roweis, Z.~Ghahramani, A unifying review of linear \mbox{Gaussian}
  models, Neural Comp. 11 (1999) 305.
\newblock \href {http://dx.doi.org/10.1162/089976699300016674}
  {\path{doi:10.1162/089976699300016674}}.

\bibitem{bishop:bayesian_pca}
C.~M. Bishop, Bayesian \mbox{PCA}, in: M.~S. Kearns, S.~A. Solla, D.~A. Cohn
  (Eds.), Advances in Neural Information Processing Systems 11, MIT Press,
  1999, pp. 382--388.

\bibitem{bishop:variational_pca}
C.~M. Bishop, Variational principal components, in: ICANN99 Ninth International
  Conference on Artificial Neural Networks, 1999, pp. 509--514.
\newblock \href {http://dx.doi.org/10.1049/cp:19991160}
  {\path{doi:10.1049/cp:19991160}}.

\bibitem{RSSB:RSSB196}
M.~E. Tipping, C.~M. Bishop, Probabilistic principal component analysis, J.
  Roy. Stat. Soc.: Ser. B (Stat. Meth.) 61 (1999) 611.
\newblock \href {http://dx.doi.org/10.1111/1467-9868.00196}
  {\path{doi:10.1111/1467-9868.00196}}.

\bibitem{Balian1987229}
R.~Balian, M.~V{\'e}n{\'e}roni, Incomplete descriptions, relevant information,
  and entropy production in collision processes, Ann. Phys. 174 (1987) 229.
\newblock \href {http://dx.doi.org/10.1016/0003-4916(87)90085-6}
  {\path{doi:10.1016/0003-4916(87)90085-6}}.

\bibitem{rau:physrep}
J.~Rau, B.~M{\"u}ller, From reversible quantum microdynamics to irreversible
  quantum transport, Phys.\ Rep. 272 (1996) 1.
\newblock \href {http://dx.doi.org/10.1016/0370-1573(95)00077-1}
  {\path{doi:10.1016/0370-1573(95)00077-1}}.

\bibitem{barbour:timeless}
J.~B. Barbour, The timelessness of quantum gravity: \mbox{I.} \mbox{The}
  evidence from the classical theory, Class.\ Quant.\ Grav. 11 (1994) 2853.
\newblock \href {http://dx.doi.org/10.1088/0264-9381/11/12/005}
  {\path{doi:10.1088/0264-9381/11/12/005}}.

\bibitem{0264-9381-11-12-007}
A.~Connes, C.~Rovelli, Von \mbox{Neumann} algebra automorphisms and
  time-thermodynamics relation in generally covariant quantum theories, Class.
  Quant. Grav. 11 (1994) 2899.
\newblock \href {http://dx.doi.org/10.1088/0264-9381/11/12/007}
  {\path{doi:10.1088/0264-9381/11/12/007}}.

\end{thebibliography}

\end{document}